\documentstyle[aps]{revtex}
\def\la{{\langle}}
\def\ra{{\rangle}}
\def\vep{{\varepsilon}}
\newcommand{\beq}{\begin{equation}}
\newcommand{\eeq}{\end{equation}}
\newcommand{\beqa}{\begin{eqnarray}}
\newcommand{\eeqa}{\end{eqnarray}}

\newcommand{\wh}{\widehat}

\begin{document}

\begin{title}
{\Large\bf Space-time properties of free motion time-of-arrival 
eigenstates.}
\end{title}
\author{\large J. G. Muga$^{1,2}$, C. R. Leavens$^1$ and J. P. Palao$^{2}$} 
\address {${}^1$  Institute for Microstructural 
Sciences,
National Research Council of Canada, Ottawa, Canada K1A OR6\\ 
${}^2$  Departamento de F\'{\i}sica Fundamental
y Experimental, Universidad de La Laguna,
La Laguna, Tenerife, Spain}
\maketitle

\begin{abstract}
The properties of the time-of-arrival
operator for free motion introduced by Aharonov and Bohm and of its 
self-adjoint variants are studied.  The domains of applicability of the 
different approaches are clarified. It is shown that 
the arrival time of the eigenstates is   
not sharply defined. However, 
strongly peaked real-space (normalized) wave packets constructed 
with narrow Gaussian envelopes centred on one of the eigenstates 
provide an arbitrarily sharp arrival time.\\

\noindent{PACS number(s):03.65.-w}
\end{abstract}   
\section{INTRODUCTION}
The extension of the classical {\it arrival time} concept to quantum 
mechanics is problematic because of the absence of trajectories in the 
standard interpretation of the quantum formalism. 
The question is important since detection of particles in time-of-flight 
and coincidence experiments are 
quite common, and quantum mechanics should
be able to predict and interpret the statistics of the arrivals.
A recent review discusses the main theoretical approaches
and open questions \cite{MSP}.   
 
The customary way to specify a dynamical variable in quantum
mechanics is to define the corresponding operator,     
but Pauli showed that a 
self adjoint time operator
conjugate to the Hamiltonian $\hat{H}$ cannot be constructed if the spectrum
of $\hat{H}$ is bounded from below \cite{Pauli}, see also
\cite{All,Gia,DM}.
There are however several ways to circumvent this objection and quantize
the classical arrival time, at least in simple cases such as free motion.
Aharonov and Bohm introduced the operator \cite{AB} 
\beq\label{top}
\wh{T}=-m\bigg{(}\widehat{q}\frac{1}
{\widehat{p}}
-\frac{\hbar}{2i}\frac{1}{\widehat{p}^2}\bigg{)}=
-\frac{m}{2}\left(\widehat{q}
\frac{1}{\widehat{p}}+\frac{1}{\widehat{p}}
\widehat{q}\right)\,,
\label{to}
\eeq
by symmetrizing the classical expression for the time-of-arrival at
$X=0$ for a particle with  
position $q$ and momentum $p$ at time $t=0$.
The same result is obtained by using  
the quantization rules of Weyl, Rivier or Born-Jordan 
\cite{SMP}, but this coincidence does not justify (\ref{to})
since, a priori, no 
rule is more fundamental than any other rule. They simply
provide a number of quantum generalizations of a classical quantity; 
one (or some)
of these can be selected  
by requiring either agreement with a particular experimental procedure 
or that certain desirable conditions be satisfied, such as
properties obeyed by the classical counterpart.
It is thus essential to carefully examine the behaviour of a given
operator and of its eigenstates to ascertain their ultimate 
physical content.   
Paul and others have examined the mathematical
properties of $\wh{T}$ as a linear operator in Hilbert
space \cite{Paul,Raza,Werner,Gia}.
In section II we shall first review these properties. Then the 
relations 
between $\wh{T}$ and positive operator valued measures, or 
its self-adjoint variants are examined, and the important
question of the domains of applicability of the theories
associated with the different operators 
is clarified.  
In section III less abstract aspects are studied, such as  
the functional
form of the eigenstates in the coordinate representation and their
time dependence. Finally, the   
behaviour of normalized wave packets ``peaked''
around one of the eigenstates is discussed in section IV.

%
%
%
%
\section{TIME OPERATORS}
$\wh{T}$ is not a self-adjoint operator but it is in a sense
the next best thing, namely, a maximal symmetric operator
\cite{Werner,Buschbook,AG}.
This means that it is hermitean (or ``symmetric'' in the mathematical
literature) but it does not admit
a self-adjoint extension. Indeed a first way to circumvent Pauli's 
argument is to relax the standard requirement and admit
hermitean but non self-adjoint operators as representations of
physical properties. By imposing hermiticity and square integrability
of the states in its range, its
domain $D(\wh{T})$ is given by the set of states with 
momentum representation obeying \cite{Paul}
\beq\label{3/2}
\lim_{p\to 0}\frac{\phi(p)}{p^{3/2}}\to 0\,.
\eeq
The eigenvalue equation in momentum representation,
$\la p|\wh{T}|T\ra=T\la p|T\ra$,
\beq
-mi\hbar\left(\frac{1}{p}\frac{d}{dp}-\frac{1}{2p^2}\right)
\la p|\psi\ra=T\la p|\psi\ra\,,
\eeq
is solved (for $p\ne$ 0) by 
\beq\label{eigen}
\la p|T,\alpha\ra=(|p|/mh)^{1/2}e^{ip^2T/(2m\hbar)}\Theta(\alpha p)\,,
\eeq
with $\alpha=\pm$.   
These (improper) eigenfunctions have the appropriate transformation
behaviour for a state arriving at 
time $T$,     
\beq\label{trans}
e^{-i\widehat{H}t/\hbar}|T,\alpha\ra=|T-t,\alpha\ra\,,
\eeq
namely, if the original state had already evolved for a 
time $t$ then it would arrive at $T-t$ instead; 
but they are not square integrable functions so they do not
define physically realizable  
state vectors in the Hilbert space 
${\cal H}=L^2$.
The constant factor in (\ref{eigen}) is
chosen to satisfy the resolution of the identity
\beq\label{one}
1_{\rm op}=\sum_\alpha\int_{-\infty}^{\infty}
dT\, |T,\alpha\ra\la T,\alpha|\,.
\eeq
In the physical interpretation of this expression, one assumes that 
all particles arrive sooner or later. 
Notice   
the inclusion of negative times. In an experimental context, all measured
arrival times are positive if the single particle state $\psi(t)$
of interest is prepared at the instant $t=0$. However,
as discussed in considerable detail by Grot et al. \cite{GRT},
the theoretical analysis is much simpler if one considers a different
problem in which it is imagined that the state is prepared at $t=-\infty$
in the state which (in the absence of interaction) evolves to the
actually prepared state $\psi(t=0)$, and the arrival time distribution
includes both negative as well as positive arrival times. This is 
very useful from
the mathematical point of view (e.g. see Eq. (\ref{one}),
or Eqs. (\ref{symx0}) and (\ref{tau})
below.)

The eigenfunctions $|T,\alpha\ra$ are complete, 
but they are not orthogonal, 
\beq
\label{overlap}
\la T',\alpha'|T,\alpha\ra=\frac{\delta_{\alpha,\alpha'}}{2}
\bigg[\delta(T-T')+\frac{i}{\pi}
{\cal P}\left(\frac{1}{T-T'}\right)\bigg]\,. 
\eeq
The non orthogonality of the eigenstates has been
associated with the 
intrinsically ``unsharp'' character of
this ``observable'' by the proponents of
the ``operational approach to 
quantum mechanics'' \cite{Buschbook}.  
These authors, and  
Gianitrappani in particular \cite{Gia},  point out that the
states $|T,\alpha\ra$  provide a ``positive operator valued
measure'' (POVM) for the arrival time. This means that for an interval of 
time $[T_1,T_2]$ the positive, 
bounded operator 
\beq\label{POVM}
\widehat{B}(T_2,T_1)=\sum_\alpha\int_{T_1}^{T_2} |T,\alpha\ra\,
dT\,\la T,\alpha| 
\eeq
can be constructed. This is not a projector, but one does not 
necessarily require projectors in order to introduce probabilities 
in quantum mechanics \cite{Buschbook,Toller}.   
By taking the trace with the normalized physical density 
operator $\wh{\rho}$,
the function 
\beq
P(T_2,T_1)=tr[\wh{B}(T_2,T_1)\wh{\rho}]\,
\eeq
fulfills in principle the conditions of a probability for arrival 
between $T_1$ and $T_2$: it is positive, 
additive for disjoint sets, and, using (\ref{one}),
tends to 1 as $T_1\to-\infty$ and $T_2\to\infty$. 
In particular, an arrival time distribution  (for a pure state 
$\psi$ at time $t$) is defined by  
\beq
\label{distri}
\Pi(T;\psi(t))=\sum_{\alpha} |\la T,\alpha|\psi(t)\ra|^2\,.    
\eeq

According to (\ref{trans}), the POVM satisfies
the ``covariance condition''
\beq\label{cc}
e^{-i\widehat{H}t/\hbar}\widehat{B}(T_2,T_1)e^{i\widehat{H}t/\hbar}
=\widehat{B}(T_2-t,T_1-t)\,.
\eeq
and this means in particular that
\beq\label{cova}
\Pi(T;\psi(0))=\Pi(T-t;\psi(t))\,,
\eeq
In words, 
the probability for arriving at $T$ is equal to the probability
for arriving at $T-t$ 
when the original state has evolved a time $t$.
This is a basic physical requirement that any
candidate for an arrival time distribution should obey.  
This is of course also true in classical mechanics.

The time-of-arrival operator $\wh{T}$ is the ``first moment''
of the POV measure. For any $\phi\in\cal H$ and $\psi\in D(\wh{T})$,
\beq\label{tb}
\la \phi|\widehat{B}^{(1)}|\psi\ra
\equiv\int T \la\phi|\widehat{B}(dT)|\psi\ra
=\la \phi|\wh{T}|\psi\ra=
\sum_\alpha\int_{-\infty}^{\infty}
dT\,\la \phi |T,\alpha\ra T \la T,\alpha|\psi\ra\,.
\eeq
There are in principle other POV measures compatible with
(\ref{cc}) and (\ref{tb}), but (\ref{POVM}) has the
following unique feature:   
Its second moment operator obeys, for states in $D(\wh{T})$,
\beq\label{var}
\Delta(\psi)\equiv \int T^2 \la\psi|\widehat{B}(dT)|\psi\ra-
\la\wh{T}\psi|\wh{T}\psi\ra=0\,,
\eeq
which is by no means an obvious relation since the eigenvectors of
$\wh{T}$ are not orthogonal (An equation like (\ref{var}) 
would be satisfied trivially by a self-adjoint operator since in that case 
the moments of the distribution of the observable are obtained as the
expectation values of the powers of the operator.)
$\Delta(\psi)$ is called the
``variance form'' \cite{Werner}.
In classical mechanics the true arrival time
distribution makes the analogous quantity minimal so this 
property has also been invoked to select 
a proper quantum distribution \cite{Ki,Werner}.

Another condition satisfied by $\Pi(T)$ is that, for the state $\psi_1$ 
defined by $\la p|\psi_1\ra=\la p|\psi\ra^*$,  
\beq\label{symx0}
\Pi(T,\psi(0))=\Pi(-T,\psi_1(0))\,,
\eeq
which follows immediately from the 
symmetry property 
\beq\label{pt}
\la p|T,\alpha\ra^*=\la p|-T,\alpha\ra\,.
\eeq
Eq. (\ref{symx0}) is also a classically motivated relation
that must hold when the arrival 
point is $X=0$ \cite{Ki,Werner}. (If only covariance and
minimum variance were imposed the arrival point would not be 
specified.)  
 
The distribution of arrival times (\ref{distri}),
was first obtained by Kijowski for states with positive momenta
or negative momenta only ($\alpha=1$ or $\alpha=-1$ but not both)
\cite{Ki}, 
and has been later rederived, studied or generalized by several authors
\cite{Werner,Busch,GRT,DM,Leon,D}. Werner in particular \cite{Werner},
with  the same restriction imposed by Kijowski, justified the 
uniqueness of (\ref{distri}) subject to the conditions
(\ref{cova}), (\ref{var}), and (\ref{symx0}). Grot, Rovelli and Tate
\cite{GRT} 
introduced a regularized self-adjoint operator and considered  
the full expression (\ref{distri}) for possible application 
to more general states having both positive and negative momenta 
but vanishing in the proximity
of $p=0$. Delgado and Muga \cite{DM}, 
with the restriction imposed by Kijowski and Werner (i.e., for states with
purely positive or negative momenta) arrive at the
distribution using a different self-adjoint operator.         
We shall next discuss these two proposals  
for self-adjoint arrival time operators.

\subsection{SELF-ADJOINT VARIANTS OF $\wh{T}$}
Grot, Rovelli and Tate \cite{GRT} trace the
non-orthogonality of the eigenstates to the singularity at $p=0$, and 
produce a ``regularized'' self adjoint time operator
$\wh{T}_\vep$
with eigenstates that differ from (\ref{eigen}) only in a small momentum
region around $p=0$,  
\beq\label{GRTs}
\la p|T\pm\ra_\varepsilon=\Theta(\pm p)(h m f_\varepsilon(p))^{-1/2}
\exp\bigg(\frac{iT}{m\hbar}\int_{\pm\varepsilon}
^p\frac{dp'}{f_\varepsilon(p')}\bigg)\,,
\eeq
where $\vep$ is a small positive number, 
\beq
f_{\varepsilon}=\cases{
1/p\,, & $|p|>\varepsilon$\cr
\varepsilon^{-2}p\,, &$|p|<\varepsilon$\cr}\,,
\eeq
and $\Theta$ is the Heaviside ``step'' function.
The integral in (\ref{GRTs}), that plays the role of the energy 
when divided by $m$, becomes
\beq
\int_{\pm\varepsilon}
^p\frac{dp'}{f_\varepsilon(p')}=\cases{
(p^2-\varepsilon^2)/2\,, & $|p|>\varepsilon$\cr
\varepsilon^{2}\ln(|p|/\varepsilon)\,, &$|p|<\varepsilon$\cr}\,,
\eeq
so that the 
regularization amounts to changing the energy spectrum in the region
around $p=0$ by introducing negative energies.
In this manner Pauli's objection is avoided. While Grot, Rovelli and 
Tate modify the eigenstates (and therefore the time operator) in
the proximity of $p=0$,
Paul discussed a similar idea \cite{Paul},
namely, to modify the physical state vector wave functions
infinitesimally around this point so that
$\wh{T}$ can be applied to them. 
He concluded however that such modifications were not physically 
meaningful since these 
infinitesimal changes may lead, for example, to arbitrary values  
of the expectation value of the square of $\wh{T}$, 
$\la \wh{T}^2 \ra$.
Note that the small $p$ region is responsible for
the long time asymptotic behaviour of the arrival time
distribution (see the Appendix A) and therefore any
infinitesimal change there
affects drastically
quantities such as $\la \wh{T}^n\ra$,
even though it will affect only infinitesimally other 
expectation values, for example
$\la \widehat{q}^n\ra$, or $\la \widehat{p}^n\ra$, 
with $n=1,2,3...$ \cite{Paul}.
Paul's observation is correct, but we shall argue in the discussion that 
the consequences are not necessarily as negative as he thought.          

A second proposal that was already pointed out by Kijowski
(section 8 of \cite{Ki})
and has been developed  further by Delgado and Muga \cite{DM} is to
take the two pieces of $\wh{T}$, 
\beq
\wh{T}=\wh{T}\Theta(\wh{p})+\wh{T}\Theta(-\wh{p})\,,
\eeq
the first  acting on the positive momentum subspace and the
second acting on the negative momentum subspace, 
and combining them with a negative sign instead,
\beq
\wh{T}_{-}=\wh{T}\Theta(\wh{p})-\wh{T}\Theta(-\wh{p})\,.
\eeq
This operator is self-adjoint, and has eigenstates $|T\ra$ formed by 
the combination 
\beq\label{tau}
|T\ra\equiv |T+\ra+|-T-\ra\,. 
\eeq
This time operator avoids Pauli's argument in a different manner. 
It is not conjugate to $\widehat{H}$ but to an operator
$\widehat{\cal H}={\rm sgn}(\widehat{p})\widehat{H}$ related to
$\widehat{H}$ by 
the change of the sign for negative momenta.
Technically, because $\wh{T}$  is a maximal symmetric operator, 
$\wh{T}_-$ cannot be its self-adjoint
extension, and only gives the same result as $\wh{T}$ when acting 
on the subspace of positive momenta (there is a change of sign
for the negative momentum subspace).   
   
So one can in principle construct self-adjoint variants of $\wh{T}$. 
But what do they mean? and what new information can we extract from them?   
Unfortunately, 
neither the states $|T\pm\ra_\varepsilon$ 
nor the states $|T\ra$ transform according to
Eq. (\ref{trans}). Equivalently, none of the self-adjoint operators 
discussed satisfies the covariance condition, and the 
``arrival time distributions'' computed with them
do not in general satisfy the basic
physical requirement (\ref{cova}).   
To avoid this problem, in both approaches the domains of {\it physical
applicability} of the self-adjoint operators have to be restricted
with respect to the mathematical domains.  
For the approach by Grot, Rovelli and Tate  
the domain should be restricted to states with momentum support 
outside the regularization region around $p=0$,
$(-\varepsilon, \varepsilon)$,
so that in fact the arrival time distribution is 
again given by (\ref{distri}). 
Similarly, the approach associated with $\wh{T}_-$  
is only physically meaningful for states with purely
positive/negative momenta 
and the corresponding distributions are therefore contained in  
(\ref{distri}). In summary, the apparent advantage of these
two proposals is misleading, since in practice they are
only applicable when their results are equivalent to 
the ones provided by the POVM  
related to the Aharonov-Bohm operator. 
In fact, the limitations on the domains of applicability of these two 
approaches are quite severe. The arrivals at a screen or detector 
will occur for all states in
$\cal{H}$ and not just for an especial set of states. A complete theory 
should provide the arrival time distribution in all cases.
The important point is that the
distribution associated with the POV measure applies for all states in 
$\cal{H}$ (irrespectively of their behaviour at $p=0$), so it is in this
sense a more complete approach. This is perfectly 
compatible with $D(\wh{T})\ne {\cal H}$ since the time operator
is only one of its moments. 
Let us recall that a probability
distribution exists independently of the existence of its moments. 
In fact in classical mechanics ensembles with non-zero probability at 
$p=0$ have no finite average arrival time, but the distribution is
nevertheless well defined. In this respect only the POVM approach provides a 
correct classical limit.   

In summary, the distribution $\Pi(T)$ is satisfactory in many
ways. There is however an important point that 
has not been considered yet. How do the eigenstates of $\wh{T}$ 
behave? Do they 
really represent states that arrive at a given time for a given
position?        
The meaning of these eigenstates, although central, 
has not been sufficiently discussed.
In particular, they
have always been studied in momentum representation without paying
attention to its coordinate representation and time dependence.
These aspects are examined in the next section.
Before doing so, it is noted that , because of the symmetry (\ref{pt}), 
the coordinate representations of the time evolved states
$|T_t\pm\ra\equiv\exp(-i\widehat{H}t/\hbar)|T\pm\ra$
are related by    
\beq\label{sym1}
\la x|T_t-\ra=\la -x|T_t+\ra\,.
\eeq
so that it is enough to study one of the cases. Further simplification 
comes from the fact that,  
as a consequence of (\ref{trans}), studying the time dependence 
of one of the states, say  $|T'_t+\ra$, from $t=0$ to the nominal
arrival time $t=T'$,
is equivalent to considering the sequence of eigenstates $|T+\ra$
from 
$T=T'$ to $T=0$.
The space-time analysis that we shall carry out in the next section
should be taken with
some precaution because 
these are not square integrable states. 
As occurs with the continuum stationary states used in
scattering theory, their physical interpretation requires the 
construction of normalizable wave packets peaked at one of 
them.
\section{COORDINATE REPRESENTATION OF THE EIGENSTATES}
The coordinate representation of the wavefunction (\ref{eigen})
is given by the integral, 
\beq\label{xinte}
\la x|T+ \ra=\int_0^\infty
\la x|p\ra\left(\frac{p}{mh}\right)^{1/2}
e^{ip^2T/(2m\hbar)} dp\,,
\eeq
where delta function normalization, $\la x|p\ra=h^{-1/2}
\exp(ixp/\hbar)$, is used for the plane waves. Let us assume for the
time being that $T>0$. The original path of integration is very
inefficient numerically because of the rapid oscillations of the
exponentials. The easiest way to calculate the integral is to deform
the contour in the complex $p$-plane along the imaginary axis from the
origin up to the intersection with the steepest descent path defined by     
\beq
p_I=mx/T+p_R\,,
\eeq
($p_R$ and $p_I$ are, respectively, the real and imaginary parts of $p$)
and then to follow this steepest descent path rightwards to 
infinity in the first quadrant.
The saddle is on the real axis at $-mx/T$.
Two cases have to be distinguished:    
For $x>0$ the path does not cross the 
saddle and the value of the integral is small. In this case the 
origin is the only relevant critical point.  
Use of Watson's lemma \cite{asym} for large $x$ and small $T$
provides the leading term  
\beq\label{xpo}
\la x|T+\ra\sim\frac{h^{1/2}e^{3i\pi/4}}{x^{3/2}\pi 2^{5/2}
m^{1/2}}\,,\:\:\:x\to\infty\,,  
\eeq
note that this asymptotic behaviour is independent of $T$.  
But for $x<0$ the saddle becomes the dominant critical point. Retaining 
the leading term \cite{asym}, 
\beq\label{xne}
\la x|T+\ra\sim\frac{(m|x|/h)^{1/2}}{T}e^{i\pi/4}
e^{-\frac{ix^2m}{2\hbar T}}\,,\:\:\: x\to-\infty\,.
\eeq
In the first case 
$\la x|T+\ra$ 
decreases as $x^{-3/2}$, whereas in the second one it increases as 
$|x|^{1/2}$ asymptotically.
In fact the integral can be expressed exactly in terms of
parabolic cylinder 
functions, but the approximate critical point treatment
just outlined is worthwhile since it
allows a simple rationalization of the exact results. 
By means of the change of variable  
\beq
s=\alpha p\,,
\eeq
where
\beq
\alpha=-\left[\frac{T}{m\hbar}\right]^{1/2}e^{-i\pi/4}\,,
\eeq
(\ref{xinte}) becomes
\beq\label{in}
\frac{1}{hm^{1/2}(\alpha^{1/2})^3}\int_C ds\, s^{1/2} 
e^{-s^2/2+zs}\,,
\eeq  
where all square roots are calculated with a branch cut on
the negative real axis, and   
\beq
z=xe^{-i\pi/4}\left(\frac{m}{\hbar T}\right)^{1/2}.
\eeq 
The integration path $C$ in (\ref{in}) goes now from $0$ to $\infty$
along the bisector of the second quadrant in the complex $s$ plane.
Note that $z$ is the saddle point of $\exp(w)\equiv\exp(-s^2/2+sz)$. It  
lies on the bisector of the second/fourth quadrant for
positive/negative $x$. The steepest descent paths from it 
are parallel to the real axis, and
$C$ lies on the border between ``hill'' and ``valley''
($w_R\equiv {\rm Real}(w)=0$). It can however be deformed 
into a line just above the branch cut since there is no contribution
at $\infty$. As the integrand across the branch cut simply
changes sign, (\ref{in}) is one half of the loop integral
around the cut.
In this fashion one of the integral forms of the parabolic
cylinder function $D_{-3/2}(z)$ can be recognized \cite{AS},
\beq\label{xt}
\la x|T+\ra=\frac{\Gamma(3/2)}
{hm^{1/2}(\alpha^{1/2})^3}e^{z^2/4}D_{-3/2}(z)\,.
\eeq
The asymptotic behaviour
for large $|z|$, see \cite{Grand}, is in agreement with the expressions 
(\ref{xpo}) and (\ref{xne}). 
The corresponding results for the case $T<0$ are simply
obtained by using the symmetry
\beq\label{sym2}
\la x|T +\ra^*=\la -x|-T+\ra\,.
\eeq
The 
asymptotic behaviour for negative and positive $x$ (growing and decaying
respectively) changes abruptly 
at $T=0$. Since $|T+\ra$  does not represent a physical state
vector, the discontinuity at $T=0$ is not problematic, but indicates again
that a literal physical interpretation of these states is not allowed.
We shall see later that normalized wavepackets formed with these 
states do not present this singularity.

There is a region close to $x=0$ that cannot be described by 
the asymptotic formulae for large argument. In that region, however,
the parabolic cylinder
function can be expressed by means of a power series
\cite{AS},
\beq\label{x0}
D_{-3/2}(z)=\frac{\Gamma(-1/4)}{\pi^{1/2} 2^{5/4}}
-\frac{\Gamma(1/4)}{\pi^{1/2} 2^{3/4}} z + {\cal O}(z^2)\,.
\eeq
Combining (\ref{x0}) and (\ref{xt}) explicit expressions
for $\la x=0|T+\ra$ and for its ``flux'' $J(x=0)$ can be obtained
\cite{note}. In particular,  
\beq
J(x=0)=\frac{[\Gamma(3/2)\Gamma(-1/4)]^2}
{(2\pi)^3 T^2 2^{3/2}}\,.
\eeq
For $T>0$, neither the wave function $\la x|T+\ra$ nor the flux
are zero at $x=0$. 
Both the ``density'' $|\la x=0|T+\ra|^2$ and
$J(x=0)$ grow monotonically
as $T\to 0$.         
Figures 1a and 1b illustrate all the dependences discussed. 
$|\la x|T+\ra|^2$ is depicted for a
series of decreasing times $T>0$ as a function of $x$
for two different scales:
between $x=-2$ and $x=0.2$ (1a), and between $x=-0.2$ and $x=0.2$ 
(1b) (atomic units are used in all numerical examples). 
In the larger scale the density of the eigenstate 
is essentially a straight line, pivoting at $x=0$,
that approaches the vertical 
as $T\to 0$. The finer scale however shows that 
the arrival is not sharply defined. 
Even though, in a loose sense, ``most of the wave'' $\la x |T+\ra$
passes from $x<0$ to $x>0$ at 
$T=0$, there is a tail at $x>0$ present for an arbitrary $T$.    
Since these states are not normalized to one it is not possible to
quantify the fraction of particles that can be found to the
right of $x=0$ before $T=0$.

One might think that the eigenstates of the self-adjoint operators, which are 
orthogonal, could avoid this 
type of unsharpness \cite{sharp}. But they don't.
For arbitrary values of $\epsilon$ we have numerically checked  
(see the Appendix B) that there is a non vanishing 
density  $|\la x|T+\ra_\varepsilon|^2$ at $x>0$ 
for fixed $T$.  
Similarly, the states $\la x|T\ra$ are also non
zero on the right hand side. Using their defining equation (\ref{tau}),  
(\ref{sym1}) and (\ref{sym2}), one finds 
\beq
\la x|T\ra=2{\rm Re} \la x|T +\ra\,.
\eeq
(It is to be noted that the corresponding time evolved state does
not take this form. 
Instead, $\la x|T_t\ra\equiv\la x|e^{-iHt/\hbar}|T\ra=
\la x|(T-t)+\ra+\la x|(T+t)+\ra^*$.)
Figure 2 represents the square of this quantity for a relatively large 
spatial interval (the small, but  non-zero density for
$x>0$ is not seen in this scale).
The figure corresponding to the imaginary part is very similar.
Note the increasingly 
rapid oscillation as $x\to -\infty$. Cancellation of these oscillations 
occurs in linear combinations of the states 
$\la x|T\ra$ or $\la x|T+\ra$ over a  
nonzero band leading to localization in a region 
closer to the origin, see Figure 3 and the discussion in the following
section.  
There is a helpful classical association to understand the oscillatory 
pattern: high velocity particles (fast oscillations) 
have to start at longer distances 
and low velocity particles (slow oscillations) at shorter distances
if they all have to arrive 
at $X=0$ simultaneously. 
\section{COORDINATE AND TIME DEPENDENCE OF THE 
NORMALIZED QUASI-EIGENSTATES}       
We shall here construct normalized wavepackets 
by using a Gaussian distribution of the states 
$\la x|T'_t +\ra$ peaked at $T'=T$ \cite{Paulfn}.
\beq\label{norma}
\la x|\Psi(t;T,\Delta T)\ra=
N\int_{-\infty}^{\infty}
e^{-\frac{(T-T')^2}{2(\Delta T)^2}}\la x|T'_t +\ra\,dT'\,.
\eeq
Carrying out the Gaussian integral over $T'$ and then determining the 
normalization constant $N$ by evaluation of 
$\int_{-\infty}^{\infty} dx\,|\la x|\Psi(t;T,\Delta T)\ra|^2$ gives the 
normalized wave packet 
\beq\label{coor}
\la x|\Psi(t;T,\Delta T)\ra=
\frac{2\pi^{1/4}(\Delta T)^{1/2}}{hm^{1/2}}
\int_0^\infty p^{1/2} e^{ipx/\hbar} e^{i(T-t)p^2/2m\hbar}
e^{-(\Delta T)^2p^4/8m^2\hbar^2}dp\,.
\eeq
This integral is readily evaluated numerically because of the 
exponential dependence on $-p^4$. The momentum representation is 
given explicitly by  
\beq\label{momre}
\la p|\Psi(t;T,\Delta T)\ra=\frac{2\pi^{1/4}(\Delta T)^{1/2}}
{(hm)^{1/2}}p^{1/2}e^{i(T-t)p^2/(2m\hbar)}
e^{-(\Delta T)^2p^4/(8m^2\hbar^2)}\Theta(p)\,.
\eeq
As $\Delta T\to 0$ these states are orthogonal
and satisfy the eigenvalue equation to any desired degree
of accuracy: 
\beqa\label{overl}
\la\Psi(t;T',\Delta T)|\Psi(t;T,\Delta T)\ra&=&w\left(
\frac{T-T'}{2\Delta T}\right)=\cases{
1 &${\rm if}\; T=T'$\cr
\sim {\cal{O}}\left(\frac{\Delta T}{T-T'}\right)& ${\rm as}\;
\frac{\Delta T}{T-T'}\to 0$\cr}\\
\la p|\wh{T}|\Psi(0;T,\Delta T)\ra&=&T\la p|\Psi(0;T,\Delta T)\ra
+{\cal O} (\Delta T)^{5/2}\,,
\eeqa 
where $w(z)=\exp(-z^2){\rm erfc}(-iz)$ is the ``$w$-function'' \cite{AS}.
The moments $\la \widehat{p}^n \ra$ of the momentum
distribution can be obtained from (\ref{momre}),
\beq
\la \widehat{p}^n\ra=\Gamma\left(\frac{n+2}{4}\right) \pi^{-\frac{1+n}{2}}
\left(\frac{\Delta T}{hm}\right)^{-n/2}\,.
\eeq
Both the average energy $\la E\ra\equiv\la \wh{p}^2\ra/2m=
h/(2\pi^{3/2}\Delta T)$ and the variance
$\Delta E=[2^{-1}-\pi^{-1}]^{1/2}
\hbar/\Delta T$ diverge as $\Delta T$ goes to zero.

Similarly, from (\ref{coor}) one finds 
\beq
\la \widehat{x}\ra
=-\frac{\Gamma(3/4)}{\pi}
\left(\frac{h}{m\Delta T}\right)^{1/2}(T-t)\,,
\eeq
whereas higher moments diverge. (A finite ``width'' can however be
defined as the half width at half height.)  
The average velocity $\la \widehat{p}/m\ra$, which is also the
velocity of the centroid
$\la \widehat{x}\ra$ is given by $\Gamma(3/4)[h/(\Delta T m
\pi^2)]^{1/2}$. The behaviour of these states is arbitrarily close to what
one could desire for an ideal arrival time eigenvector: They are normalized,
and there is no discontinuity at $t=T$;
for a fixed $\Delta T$, 
they still obey a transformation law of the form (\ref{trans});    
the wave travels towards the origin with constant velocity, and the
average position crosses the origin at time $t=T$, which is also the time 
when the spatial width of the wave packet attains its minimum value. 
Of course a certain 
unsharpness remains: From 
(\ref{coor}) one finds 
$\la -x|\Psi(2T-t;T,\Delta T)\ra=\la\Psi(t;T,\Delta T)|x\ra$, so the
probability density has inversion symmetry in $(x,t)$ with respect to 
the space time ``origin'' $(0,T)$.
In particular, for $t=T$, half the norm is
to the right of $x=0$ independently of $\Delta T$
(The question of the existence of quantum states  
where the particle stays strictly  
on one side of $X$ before $T$ and on the other side after $T$
is addressed in the Appendix C.)  
However, the 
wave packet density and flux are  
more and more peaked at the space-time point ($x=0,\, t=T$)
as $\Delta T \to 0$, so that     
the passage of probability from
left to right is sharp to any desired accuracy,
\beqa
|\la x=0|\Psi(t=T;T,\Delta T)\ra|^2&=&
\left(\frac{m}{h\Delta T}\right)^{1/2}\frac{\Gamma(3/8)^2}
{\pi 2^{5/4}}\\
J(x=0,t=T)&=&\frac{\Gamma(3/8)\Gamma(5/8)}{\Delta T \pi^{3/2}
2^{3/4}}\,.
\eeqa
The normalized wavepackets constructed with the eigenstates (\ref{GRTs})  
have been examined in \cite{Jonathan}, where it is reported that this 
arbitraryly sharp accuracy is not found in that case. 

%
%
%
%
%
\section{Discussion}
Given the importance of the timed detection of particles 
at screens or in time-of-flight experiments, finding a theoretical 
description of the arrival times seems imperative.
Apparatus dependent results are available, for example via modelling 
the detection with
complex absorbing potentials or other measurement
models \cite{AP,Aetall,Halli}, but it
is reasonable to inquire 
if an intrinsic, apparatus-independent distribution can be 
naturally defined by means of the usual operator approach to 
quantum mechanics. We have seen in sections II and III that the POVM
and the corresponding distribution $\Pi(T)$ associated with the time
operator $\wh{T}$ provide a rather satisfactory answer from the 
point of view of the properties satisfied: covariance with respect to 
time translations, minimum variance, appropriate symmetries, 
physically correct  
domain of applicability, and sharp space-time behaviour of the 
normalized quasi-eigenstates.

One of the objections by Paul to $\wh{T}$, 
which is also a shortcomming of the theories based on self-adjoint operators, 
namely the restrictive domain of the time operator(s) (which does not include,
for example, states such as minimum uncertainty product Gaussians) 
is overcome by the POVM theory.       
An important point is to consider the primary object as the POVM
(equivalently the resolution of the identity, or the arrival time 
distribution) rather than the operator.
In this manner, the domain of applicability of the theory 
is $\cal{H}$, and the classical limit is correct.
Moreover in this light the other problem indicated by
Paul, namely the 
extreme sensibility of the expectation values of powers of $\wh{T}$
to small perturbations, is relatively unimportant. 
It is a fact that some quantities are very sensitive to certain small
perturbations and we are simply  dealing with one of them. The moments 
in the classical case would also suffer from such a sensitivity.
The moral is that one should not pay as much attention to the moments
of a time-of-arrival distribution (highly unstable with respect to small 
perturbations or changes in the apparatus resolution, and possibly 
divergent) but to general
features of the distribution (peaks, global form, or width at half 
height for example). A consequence is that an uncertainty principle in
terms of $\la \wh{T}^2\ra$ is not of much use, since this quantity
will generally diverge. If it does not, it will be too unstable with
respect to small perturbations. It is preferable to express the 
uncertainty principle in terms of other measurements of width, such as 
the the half width at half height.               

Apart from the the positive features of $\Pi(T)$, 
it is also fair to point out several unclear points or    
open questions. 
For example, it is necessary to generalize the treatment
to higher dimensions or to scattering
problems.  
The connection between $\Pi(T)$ and measurements
(especially in non-classical cases) is also pending, see also
\cite{Aetall}.   
We can however advance as a preliminary analysis two 
important features of
a hypothetical measurement of $\Pi(T)$:  
The resolution of identity (\ref{one}) implies
the following structure of $\Pi(T)$,
\beq\label{ddd}
\Pi(T)=|\la T+|\psi_+\ra|^2+|\la T-|\psi_-\ra|^2\,,
\eeq
where $|\psi_\pm\ra=\Theta(\pm \wh{p})|\psi\ra$.
This implies ignoring the ``interference terms'' $|\psi_-\ra\la\psi_+|$
and $|\psi_+\ra\la\psi_-|$, so that in a hypothetical operational procedure to 
measure $\Pi(T)$, only the diagonal terms of the density operator 
contribute, 
\beq
\wh{\rho}\to |\psi_+\ra\la\psi_+|+|\psi_-\ra\la\psi_-|\,.
\eeq
Apart from the fact that the
practical implementation of this diagonalization may be cumbersome, 
the neglect of interferences is not a desirable feature, since    
many different quantum states would give the 
same distribution.     
A second problematic aspect is associated with the interpretation of 
$|\la T+|\psi_+\ra|^2$ as the contribution to $\Pi(T)$ from particles
arriving from the left and of $|\la T-|\psi_-\ra|^2$ as the contribution
from those arriving from the right. This is particularly evident when 
the {\it backflow effect},
\beqa
J\le 0\;\; &{\rm for}&\;\; |\psi\ra=|\psi_+\ra\,,\\
J\ge 0\;\; &{\rm for}&\;\; |\psi\ra=|\psi_-\ra\,,
\eeqa
occurs. Then
$\Pi(T)$ assigns zero probability to arrivals from the ``anomalous side'',
(e.g. from the right when $|\psi\ra=|\psi_+\ra$). 
This implies that if $|\psi\ra=|\psi+\ra$ or  $|\psi\ra=|\psi-\ra$
then either particles should be found to arrive either only from
the left or only 
from the right, respectively, even during the time interval when $J$ has the 
``wrong sign'', or that the theory is appropriate when the ``screen'' 
is ``one-sided'', failing to detect any particles arriving from the
``anomalous 
side''. Now consider the corresponding implication 
for the general state $|\psi\ra=|\psi+\ra+|\psi-\ra$ with both $|\psi+\ra$ 
and $|\psi-\ra$ nonzero. One possibility is that the interference terms 
do not in fact contribute to the intrinsic arrival time distribution.
The other is that the distribution (\ref{ddd}) is only
appropriate when the apparatus measures the sign of the momentum of each 
incident particle, thus collapsing the wavefunction of that particle either 
to $|\psi+\ra$ or $|\psi_-\ra$, and then switches on the appropriate 
one-sided detecting screen.    

\acknowledgments{
One of us, J. G. M., acknowledges a very useful correspondence with 
Marco Toller. The content of this paper was presented as a talk in 
the workshop ``Time in quantum mechanics'', La Laguna, May 1998.
The work has been supported by Gobierno Aut\'onomo
de Canarias (Spain)
(Grant No. PB2/95), and by CERION.}  

\appendix    
\section*{A: LARGE  T BEHAVIOUR  OF $\Pi(T)$}

Let us consider the asymptotic, large $T$ behaviour of 
$\la T+|\psi\ra$,
\beq
\la T+|\psi\ra=\frac{1}{(mh)^{1/2}}\int_0^\infty
p^{1/2} e^{-ip^2T/(2m\hbar)}\la p|\psi\ra dp\,.
\eeq
Assumming that $\la p|\psi\ra$ can be analytically continued into the 
fourth quadrant in the complex $p$-plane we deform the integral contour 
into the ray $p=\gamma e^{-i\pi/4}$, ($0\le\gamma<\infty$). With the change 
$g=\gamma^2/(2m\hbar)$ the integral takes the form 
\beq
\la T+|\psi\ra=\frac{(mh)^{1/4}}{2\pi^{3/4}}e^{-i\pi3/8}
\int_0^\infty dg\, e^{-gt} \psi(g) g^{1/4}\,,
\eeq
where the origin appears as the critical point. If, as $g\to 0$,
$\psi(g)\sim c g^u$ (where $u$ is not necessarily an integer), 
use of Watson's lemma provides the dominant term,
\beq
\la T+|\psi\ra\sim e^{-i\pi3/8}\frac{c(mh)^{1/4}}{2\pi^{3/4}}
\frac{\Gamma(u+5/4)}{T^{u+5/4}}\,.
\eeq
\section*{B: COORDINATE REPRESENTATION OF $|T+\ra_\vep$}
The integral over $p$ for the coordinate representation $\la x|T+\ra_\vep$
can be separated into two parts, from $0$ to $\vep$, $I_1$, and from
$\vep$ to
$\infty$, $I_2$. The second one can be obtained numerically with the 
treatment of section III, by deforming the 
contour and using the steepest descent path from the saddle. 
$I_1$ can be expressed as,      
\beq
I_1=\frac{\vep^{1-iT\vep^2/(m\hbar)}}{hm^{1/2}}
\int_0^\vep e^{ixp/\hbar}p^{\left(\frac{iT\vep^2}{m\hbar}
-\frac{1}{2}\right)}\,.
dp
\eeq
With the change of variable $u=-ixp/\hbar$ the incomplete gamma 
function is recognized, 
\beqa
I_1&=&\frac{\vep^{3/2}}{hm^{1/2}}\left(\frac{x\vep}{i\hbar}\right)^{-A}
\gamma(A,-ix\vep/\hbar)\\
&=&\frac{\vep^{3/2}}{hm^{1/2}}
\sum_{n=0}^\infty \frac{(ix\vep/\hbar)^n}{n! (A+n)}\,,
\eeqa
where $A=iT\vep^2/(m\hbar)+1/2$.
It is easy to calculate an upper bound on the $p \leq \epsilon$ contribution,
$I_1(x,t;T)$, to $<x|T_{t} +>$. The change of variable $p=\epsilon u$ gives
\begin{equation}
I_1(x,t;T)=\frac{\epsilon^{3/2}}{h m^{1/2}}
\int_{0}^{1}\frac{du}{u^{1/2}}\exp\Bigl[i\Bigl(\frac{\epsilon xu}{\hbar}
-\frac{\epsilon^{2}tu^2}{2m\hbar}+\frac{\epsilon^{2}T}{\hbar m}\ln u
\Bigr)\Bigr].
\end{equation}
Hence, both the real and imaginary parts of $I_1(x,t;T)$ are bounded in 
absolute value by
\begin{equation}
\frac{\epsilon^{3/2}}{h m^{1/2}}\int_{0}^{1}\frac{du}{u^{1/2}}
=\frac{2\epsilon^{3/2}}{h m^{1/2}}
\end{equation}
and $|I_1(x,t;T)|\leq (2\epsilon)^{3/2}/(h m^{1/2})$, independent of $x$,$t$
and $T$.

\section*{C: STRICTLY SHARP ARRIVAL STATES}
Are there quantum mechanical states where the particle stays strictly  
on one side of $X$ before $T$ and on the other side after $T$?
We have seen, by examining the coordinate representation of the eigenstates 
of different time operators that 
none of the proposals (fulfilling either the covariance condition or the 
self-adjointness condition) satisfies strictly this requirement. 
Is this an inherent  limitation of standard 
quantum mechanics? We shall argue that there are no quantum states, pure or 
mixed, that 
satisfy fully this condition.  
To this end let us use the equivalent phase space Weyl-Wigner
formalism. Its advantage 
for free motion dynamics is that the evolution kernel and dynamical equation
of motion  
are  equal in classical 
and quantum mechanics \cite{PS}. Therefore,  a
Liouville theorem applies, so that each phase
space point carries its own ``probabilistic weight''
(that can be negative in the 
quantum case), so that in the intermediate calculations one may think
and operate classically,
the only difference being in the domains of states allowed in both
mechanics and in the interpretation of the formalism \cite{PS,JCP}.
The Wigner distribution $f(q,p)$ in the position-momentum  phase space 
represents a valid quantum mechanical state if the  
associated density operator $\widehat{\rho}$ is positive,
see e.g. \cite{sing}. 
Necessary conditions may be found for $f(q,p)$ itself, such as 
\beqa
\label{con1}
&&|f(q,p)|\leq(2/h)\\
\label{con2}
&&h\int f(q,p) dq dp\leq 1\,.
\eeqa
What kind of ensemble of classical particles, having negative positions at 
time $t=0$,
would arrive 
at the same time $T$ at a point $X=0$?. Since momentum is conserved, it is
necessary a coordinated motion where faster particles start 
moving further away and slower particles start closer to $X=0$ so that they
all arrive at the same time. The phase space density 
that satisfies these requisites is     
\beq\label{ft}
f_T(x_0,p;t=0)=g(x_0)\Theta(-x_0)\delta(x_0+pT/m)\,,
\eeq
with $g(x_0)\ge 0$. 
But this distribution is too singular to satisfy (\ref{con1}) or
(\ref{con2}).

\newpage
\centerline{\large FIGURE CAPTIONS\vspace*{.2cm}\\}

{\bf Figure 1a} $|\la x|T+\ra|^2$ versus $x$
for $T=0.01$ (long dashed line), $0.005$ (short dashed line), 
and $0.001$ (solid line); $m=1$. All quantities in atomic units. 

{\bf Figure 1b} Same as Figure 1a for a smaller $x$ interval.   


{\bf Figure 2} $|\la x|T\ra|^2$ versus $x$ for $T=0.01$, $m=1$.

{\bf Figure 3} Probability density of normalized
quasi-eigenstates, see (\ref{norma}), for $T=0.04$, $m=1$, 
$\Delta T=0.002$, $t=0$ (solid line), $t=0.02$ (long dashed line),
and $t=0.04$ (short dashed line).

\begin{references}
\bibitem{note} The ``flux'' is evaluated with the standard expression, 
$\hbar/m\Im{\rm m}\{\psi^*\psi'\}$. Neither this ``flux'' nor the
``probability
density'' $|\psi|^2$ obtained from (\ref{xt}) have the proper dimensions.
This is due to the ``continuum'' normalization used for the
eigenstates of $\widehat{T}$, which are not square integrable. The
appropriate domensions are of course obtained 
by integrating these objects over time to form wave packets.
\bibitem{sharp} According to the terminology of ref. \cite{Buschbook}
any self-adjoint 
operator is ``sharp''. But refering to the eigenstates of the self-adjoint 
time operators we use the concept sharp/unsharp in
a way closer to the conventional usage of the words. Being sharp is  
``having an edge or point'', or ``being abrupt''.
\bibitem{Paulfn} A similar construction was carried out by Paul with
a flat weighting function instead of a Gaussian.
However, while he dismissed these states because they do not belong to 
$D(\wh{T})$, this is not essential to 
clarify the physical content of
the eigenstates $|T,\alpha\ra$.
\end{references}
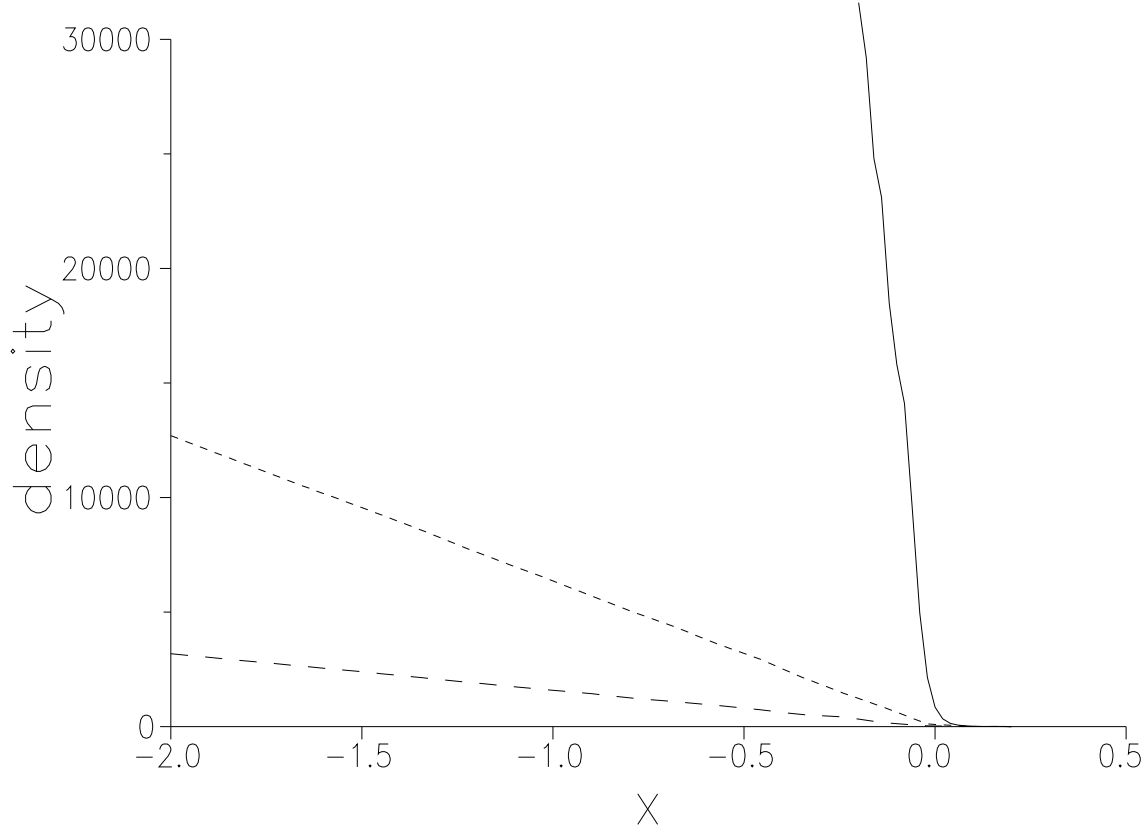
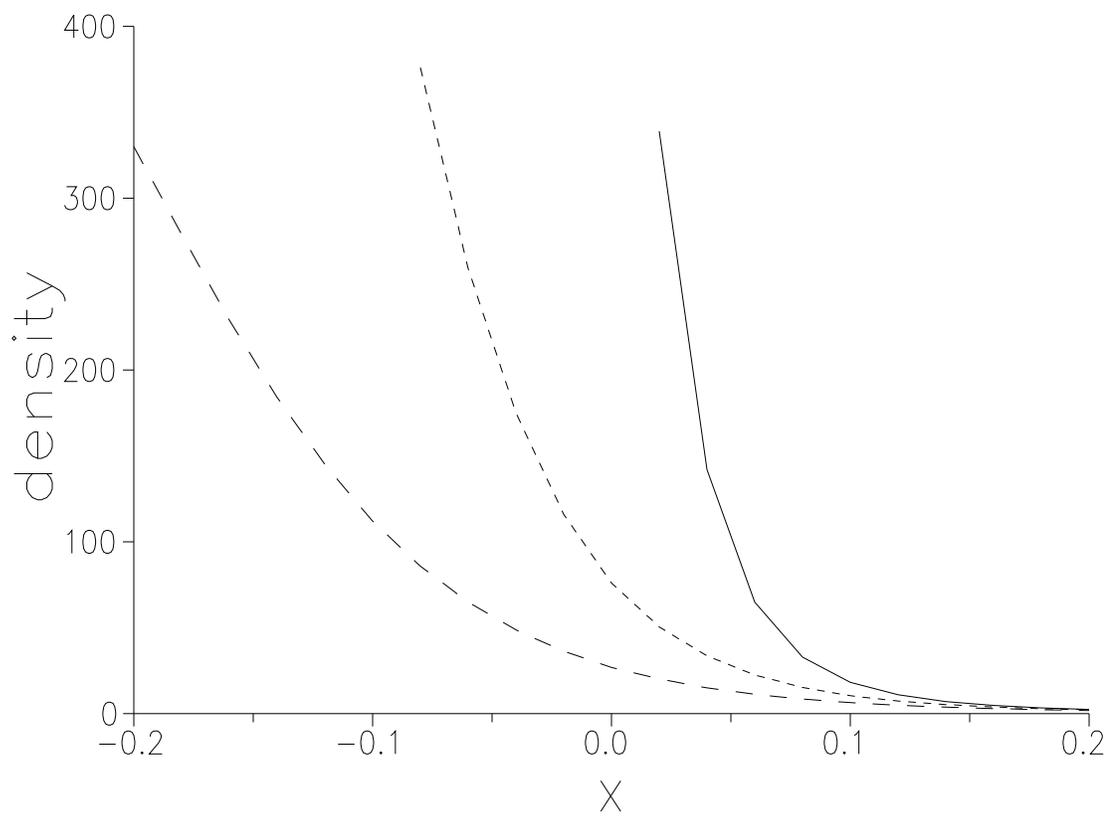
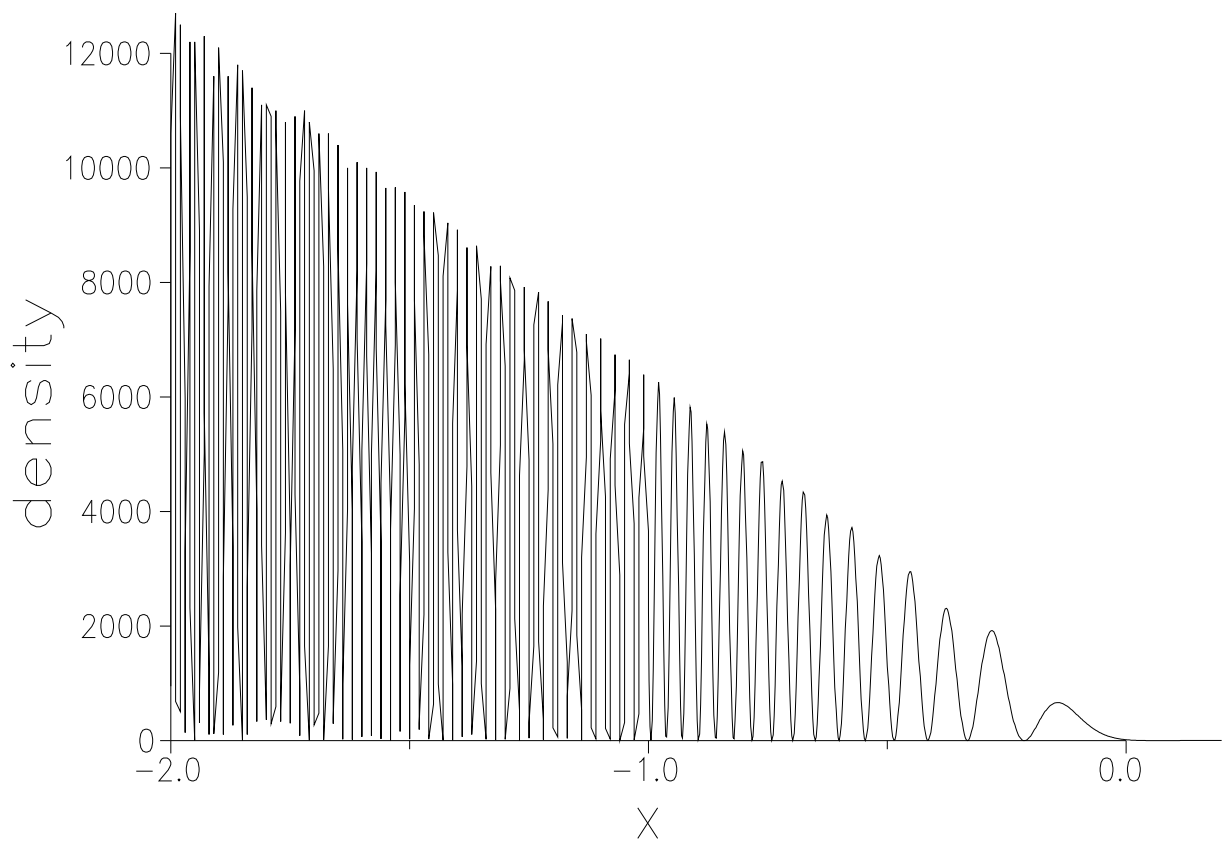
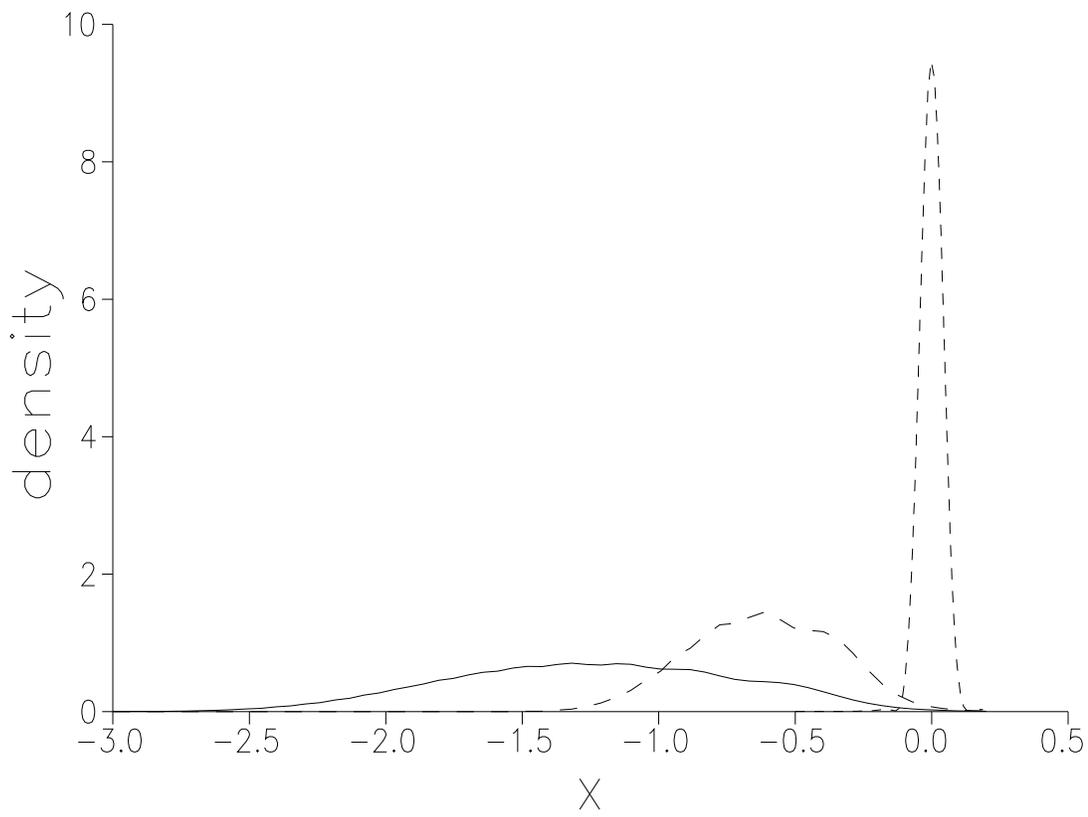
\end{document}